\documentclass[aps,amssymb]{revtex4-1}                  
\usepackage{graphicx}
\usepackage{amsmath}
\usepackage{float}
\usepackage{subfigure}
\usepackage{graphicx}
 \usepackage{mathptmx}

\newcommand{\qmean}{\overline{q_t}}

\newcommand{\qfinal}{\overline{q_\infty}}

\newcommand{\qfinalN}{\overline{q_\infty(N)}}

\newcommand{\qfinalNmax}{\overline{q_\infty(N_{\rm max})}}
\newcommand{\sizeFL}{f_t(N)}

\newcommand{\qfinalinfN}{\overline{q_\infty(N\rightarrow\infty)}}

\newcommand{\mt}{\tau}

\begin{document}
\title{Long-Time Predictability in Disordered Spin Systems Following a Deep Quench}

\author{J.~Ye}
\email{jingy@princeton.edu} 
\affiliation{
Department of Operations Research and Financial Engineering,
Princeton University,
Princeton, NJ 08544 USA}

\author{R.~Gheissari}
\email{reza@cims.nyu.edu}
\affiliation{Courant Institute of Mathematical Sciences, New York University, New York, NY 10012 USA}
 
\author{J.~Machta}
\email{machta@physics.umass.edu}
\affiliation{
Physics Department,
University of Massachusetts,
Amherst, MA 01003 USA;
Santa Fe Institute,
1399 Hyde Park Road, Santa Fe, NM 87501 USA}

\author{C.M.~Newman}
\email{newman@cims.nyu.edu}    
\affiliation{Courant Institute of Mathematical Sciences, New York University, New York, NY 10012 USA and NYU-ECNU Institute of Mathematical Sciences at
NYU Shanghai, 3663 Zhongshan Road North, Shanghai 200062, China}

\author{D.L.~Stein} 
\email{daniel.stein@nyu.edu}
\affiliation{Department of Physics and Courant Institute of Mathematical Sciences,
New York University, New York, NY 10012 USA and NYU-ECNU Institutes of Physics and Mathematical Sciences at NYU Shanghai, 3663 Zhongshan Road North, Shanghai, 200062, China}

\begin{abstract}
We study the problem of predictability, or ``nature vs.~nurture'', in several disordered Ising spin systems evolving at zero temperature from a random initial state: how much does the final state depend on the information contained in the initial state, and how much depends on the detailed history of the system? Our numerical studies of the ``dynamical order parameter'' in Edwards-Anderson Ising spin glasses and random ferromagnets indicate that the influence of the initial state decays as dimension increases. Similarly, this same order parameter for the Sherrington-Kirkpatrick infinite-range spin glass indicates that this information decays as the number of spins increases. Based on these results, we conjecture that the influence of the initial state on the final state decays to zero in finite-dimensional random-bond spin systems as dimension goes to infinity, regardless of the presence of frustration. We also study the rate at which spins ``freeze out'' to a final state as a function of dimensionality and number of spins; here the results indicate that the number of ``active'' spins at long times increases with dimension (for short-range systems) or number of spins (for infinite-range systems). We provide theoretical arguments to support these conjectures, and also study analytically several mean-field models: the random energy model, the uniform Curie-Weiss ferromagnet, and the disordered Curie-Weiss ferromagnet. We find that for these models, the information contained in the initial state does {\it not\/} decay in the thermodynamic limit--- in fact, it {\it fully determines\/} the final state. Unlike in short-range models, the presence of frustration in mean-field models dramatically alters the dynamical behavior with respect to the issue of predictability.
\end{abstract}

\maketitle

\section{Introduction}
\label{sec:intro}

The dynamical properties of Ising spin systems far from equilibrium, and in particular those following a deep quench, continue to be a major focus of research.
There are multiple directions along which this research has been pursued, including the general areas of phase-ordering kinetics~\cite{Bray94,Stauffer94,SKR118,SKR119,KKR}, persistence~\cite{DBG,J99,MS96,MR00,DHP96},
and damage spreading~\cite{Kauffman,Creutz,SSKH,Grassberger}, among many others. 
In~\cite{NS99a} another line of investigation was proposed: the problem 
of {\it predictability\/} and retention of information
in discrete spin systems evolving far from equilibrium.  

Following a deep quench an Ising spin system will be in a random configuration.
One can then ask, how rapidly does the information contained in the initial state decay over time, as a function of dimension, lattice type, model (uniform ferromagnet, random ferromagnet, spin glass, and so on), 
and other characteristics of the system under study? The nature of information 
retention or decay depends on two sources of randomness: that contained in the 
initial configuration and that generated through the dynamical realization 
governing an individual history. This dynamical randomness is present even at zero temperature, in the order that spins are chosen to attempt to flip (and in the tie-breaking rule for homogeneous systems when a spin flip would cost zero energy). Disordered Ising models such 
as random ferromagnets and spin glasses bring in a third source of 
randomness, namely in the couplings determining the local environment of an 
individual spin. We are interested in the question of how much information 
contained in the configuration at time $t$ depends on the initial configuration and random environment, and 
how much depends on the specific dynamical path the system has followed. 
We will formalize these remarks in Sect.~\ref{sec:preliminaries} below.  We have colloquially referred to the above as the ``nature vs.~nurture'' problem, where nature represents the information contained in the initial configuration and (quenched) random couplings, while nurture refers to the history (i.e., dynamical realization) of the subsequent evolution of the system. 

It was shown in early work that this problem can be solved exactly for $1D$ random
ferromagnets and spin glasses~\cite{NNS00}. Preliminary numerical studies on the $2D$~{\it homogeneous\/}
ferromagnet on the square lattice were reported in~\cite{ONSS06}. 
More recently, extensive numerical studies~\cite{YMNS13} have 
largely solved the nature vs.~nurture problem for the uniform~$2D$ ferromagnet, so that
the rate of decay of initial information for this model is now understood quantitatively. These results will be briefly reviewed in the next section.

In this paper we turn our attention to {\it disordered\/} Ising models in dimensions greater than one.
Our particular focus will be on the behavior of the random ferromagnet 
and the Edwards-Anderson~(EA) spin glass~\cite{EA75}
as a function of dimension. We will also consider the infinite-range Sherrington-Kirkpatrick~(SK) spin glass~\cite{SK75}
as a function of system size $N$. Our conclusions will be based on numerical studies, but in Sect.~\ref{sec:discussion} we
also present an analytical discussion of these models, as well as the random energy model (REM)~\cite{Derrida80} and the uniform and disordered Curie-Weiss ferromagnets.

\section{Preliminaries}
\label{sec:preliminaries}

Consider a set of Ising spins $S_i=\pm 1$ on the sites $i$ of the Euclidean lattice $\mathbb{Z}^d$. The system Hamiltonian is
\begin{equation}
\label{eq:EA}
{\cal H}=-\sum\limits_{\langle i,j \rangle}J_{ij}S_iS_j\, ,
\end{equation}
where $\langle i,j \rangle$ indicates a sum over all nearest neighbor pairs and the couplings are independent, identically distributed random variables chosen from a common distribution (which depends on the exact model under consideration).
We study numerically two types of finite-dimensional models: the first is the Edwards-Anderson~(EA) Ising spin glass~\cite{EA75} in $d$ dimensions, in which the common distribution of the couplings is a Gaussian with mean zero and variance~one.
The second is the random ferromagnet, where the couplings are all positive; here the common distribution is taken to be a 
one-sided Gaussian, in which each bond is chosen as the absolute value of a standard Gaussian random variable (again with mean zero and variance one).
Finally, we consider the infinite-range Sherrington-Kirkpatrick~(SK) spin glass. 
Here the spins sit on the $N$ sites of a complete graph, with a modified Hamiltonian
\begin{equation}
\label{eq:SK}
{\cal H}=-\frac{1}{\sqrt{N}}\sum\limits_{i<j}J_{ij}S_iS_j\, .
\end{equation}
The couplings are again chosen from a Gaussian distribution with mean zero and 
variance one, and the rescaling factor
$N^{-1/2}$ ensures a sensible thermodynamic limit of the energy and free energy per spin. 
(Note, however, that the rescaling factor plays no role in the dynamics described below.) 

Initially the spins are in a random initial configuration, in which each spin is chosen to be $+1$ or $-1$ with probability 1/2, independently of all the others. This corresponds to an infinite-temperature spin configuration. The subsequent evolution is governed by zero-temperature Glauber dynamics, which in the simulations to be described below is implemented as follows. At each step, a site $i$ is selected uniformly at random
and the energy change $\Delta E_i$ associated with flipping the associated spin is computed: $\Delta E_i=\Delta{\cal H}$ when $S_i \rightarrow -S_i$ and all other spins remain fixed.
If the energy decreases ($\Delta E_i < 0$) as a result of the flip, the flip is carried out. 
If the energy increases ($\Delta E_i > 0$), the flip is not accepted. If the energy remains the same ($\Delta E_i = 0$), 
a flip is carried out with probability 1/2. This last ``tie-breaking'' rule is relevant only for models (such as the homogenous ferromagnet, or the $\pm J$ spin glass) in which a zero-energy flip can occur. For disordered models with continuous coupling distributions, such as those considered here, this possibility never arises. These dynamics are run until the system reaches an absorbing state that is stable against all single spin flips.

In~\cite{YMNS13} the authors defined a ``heritability exponent'' by preparing
two Ising~systems (regarded as identical twins)
on a square of side $L$ with the same initial configuration and then allowing them
to evolve independently using different dynamical realizations of zero-temperature Glauber dynamics~\cite{note2d}. The spin
overlap between the resulting copies is  $q_t(L)=\frac{1}{L^2}\sum_{i=1}^{L^2}S_{i}^1(t)S_{i}^2(t)$, where $S_{i}^k(t)$
denotes the state of the $i^{\rm th}$ spin at time $t$ in twin~$k$, where $k=1,2$.  The influence of initial conditions is
quantified by $q_t(L)$, where $q_0(L)=1$ for any~$L$.  The authors examined the
average $\overline{q_t(L)}$ over both initial conditions and dynamics to investigate two relevant quantities: $\overline{q_\infty(L)} = \lim_{t \rightarrow \infty} \overline{q_t(L)}$, the size dependence of the overlap at large times,
and $\overline{q_t} =\lim_{L \rightarrow \infty} \overline{q_t(L)}$, the 
time dependence of the overlap in large volumes. 

It is important to note that heritability is not the same as
persistence, although the two ask related questions. Persistence asks which spins have not flipped up to a
time $t$, while heritability asks to what extent the
{\it information\/} contained in the initial state persists up to time
$t$. A spin may have flipped multiple times during this interval but its
final state might still be predictable knowing the initial condition. 

Our key finding for the $2D$ uniform ferromagnet~\cite{YMNS13} was that heritability decays as a power
law at long times: $\overline{q_t}\sim t^{-\theta_h}$.  The power-law exponent $\theta_h$
is the ``heritability exponent'' referred to above. The size dependence of the final overlap between twins on a finite lattice similarly
decays as a power law: $\overline{q_\infty(L)}\sim L^{-b}$, The exponents
$\theta_h$ and $b$ were shown, through a finite size scaling
ansatz, to be related by $b=2\theta_h$, consistent
with our numerically determined values and with  exact $1D$ values.

Before turning to a study of nature vs.~nurture in disordered Ising systems, we point 
out an important difference between the uniform ferromagnet on $\mathbb{Z}^d$ and
the models considered here: as noted above, there can be no ``ties'', 
or zero-energy flips in disordered models whose couplings are random 
variables arising from continuous distributions.
(This is equally true for the homogeneous ferromagnet with an odd number 
of neighbors, such as the 
$d=2$ honeycomb lattice, but
we do not consider those models here.) It was proved in~\cite{NNS00} that, in the uniform ferromagnet on the square lattice under the dynamics described here, every spin 
(on the infinite lattice) flips infinitely often. It was also proved in~\cite{NNS00} that, in any discrete spin model such as those considered here, every spin makes only a finite number
of energy-lowering flips in any dynamical run; consequently, $\overline{q_t}$ does not decay to zero as $t\to\infty$ in any finite-dimensional EA spin glass or random ferromagnet with a continuous coupling distribution~\cite{notezero}.  
In~\cite{NNS00}, we defined a dynamical order parameter $q_D$ ($D$ for dynamical) which effectively corresponds to
\begin{equation}
\label{eq:qD}
q_D=\lim_{t\to\infty}\overline{q_t}\, .
\end{equation}
Because we are now considering heritability as a function of dimension $d$, to avoid confusion we will hereafter refer to the dynamical order parameter
as $q_\infty$ rather than $q_D$. The interesting question is how does $q_\infty$ behave as a function of dimension $d$ for these models, and in particular does it tend to zero as $d\to\infty$?
For the SK~model, in contrast, one is forced to consider $q_\infty$ to be the long-time limit of $\overline{q_t(N)}$
for finite number of spins $N$, which of course will be nonzero for any finite $N$. The corresponding question is then whether $\overline{q_\infty(N)}\to 0$
as $N\to\infty$.  

\section{Models and Methods}
\label{sec:maa}

In order to distinguish the influence of nature and nurture in the above three models, 
we use the twin method described above.  As a function of time $t$, 
we look at the overlap $q$ between the twins for a system of size $N=L^d$ with periodic boundary conditions:  
\begin{equation}
q_t(N) = \frac{1}{N}\sum\limits_{i=1}^{N}S_i^1(t)S_i^2(t)
\end{equation}
where $S_i^1(t)$ denotes the state of the $i^{th}$ spin at time $t$ in twin 1 and  $S_i^2(t)$ is the spin state for twin 2. Here and throughout this work, time  $t$ is measured in sweeps, where one sweep corresponds to $N$ spin-flip attempts.
The overlap is initially unity: $q_0(N)=1$ and, on average, decays in time. For each finite realization, it reaches a final value when both twins are in absorbing states; we are therefore interested, in addition to $q_\infty$ defined above, 
in the $N$ (equivalently, $L$) dependence of $q_t(N)$ for the three models.  We also study the time to reach the absorbing state and the fraction of active or flippable spins as a function of time.  

We use the algorithm introduced in~\cite{YMNS13}.  For the first $t_0$ sweeps we implement Glauber dynamics directly: A spin $S_i$ is randomly selected,  the energy change $\Delta E_i$ is computed, and the spin is flipped ($S_i \rightarrow -S_i$) if $\Delta E_i < 0$ and not flipped otherwise.  After each attempted spin flip, time is incremented by $1/N$ sweeps.  After $t_0$ sweeps, only a few active spins (such that $\Delta E_i < 0$) remain and the dynamics is  significantly accelerated using kinetic Monte Carlo methods. To implement kinetic Monte Carlo, a list of active spins is maintained.   A single step of kinetic Monte Carlo consists of selecting a spin at random from the active list and then carrying out the flip.   The time is incremented by 1/$\sizeFL$,  where $\sizeFL$ is the number of active spins before the spin flip.  After the spin flip the active list is updated: the spin at $i$ is removed from the list and its neighbors are all checked to see if they must be added to or removed from the active list. Kinetic Monte Carlo dramatically improves the run time for the $d$-dimensional models.  For the finite-dimensional models, $t_0$ is set to 10 while for the SK model, $t_0=20$.

For each system, we study 30,000 independent pairs of twins (i.\ e.\ 60,000 systems).  From $q_t(N)$, 
we compute the mean $\qmean$, the standard deviation $\sigma_t$ and the standard error of the mean.  
We are mostly interested in the final value $\qfinal$ when both twins have reached the absorbing states.

\section{Results}
\label{sec:results}

In this section we present numerical results for the Edwards-Anderson (EA) spin glass in dimensions $d=2$, 3 and 4, the random ferromagnet in $d=2$, and then the Sherrington-Kirkpatrick (SK) model.

\subsection{Edwards-Anderson model in $d$ dimensions}
\label{subsec:EA}
\begin{figure}[H]
\centering 
\subfigure{ 
\label{fig:subfig:a}  
\includegraphics[scale=0.8]{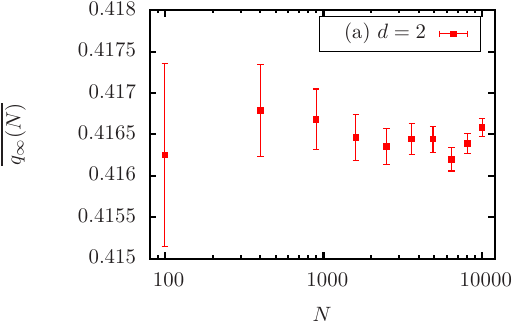}} 
 \subfigure{ 
\label{fig:subfig:b} 
\includegraphics[scale=0.8]{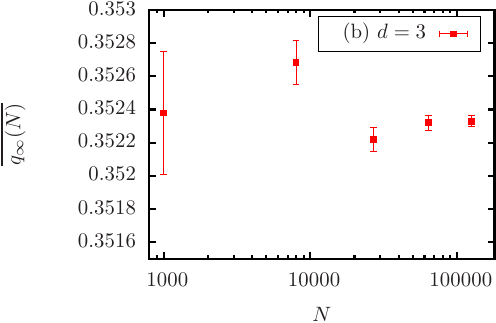}} 
\subfigure{ 
\label{fig:subfig:c} 
\includegraphics[scale=0.8]{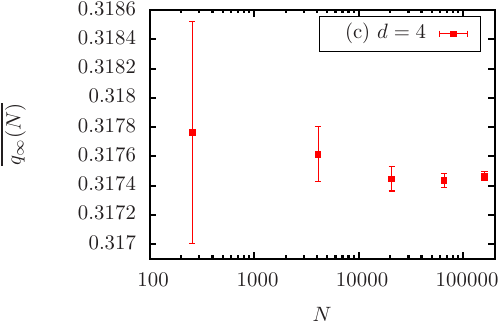}} 
\caption{The overlap $\qfinalN$ vs.\ system size $N$.  Panels (a), (b) and (c) represent Edwards-Anderson spin glasses in $d=2$, 3 and 4 dimensions, respectively.} 
\label{fig:QSD} 
\end{figure}

Figure \ref{fig:QSD} is a plot of $\qfinalN$  as a function of the number of spins $N$ for the EA spin glass, with panels (a), (b), and (c) representing $d = 2$, 3, and 4 dimensions, respectively.  It is  clear that $\qfinalN$ is rapidly converging to a non-zero constant as $N \rightarrow \infty $.  

\begin{figure}[H]
\centering  
\includegraphics[scale=1]{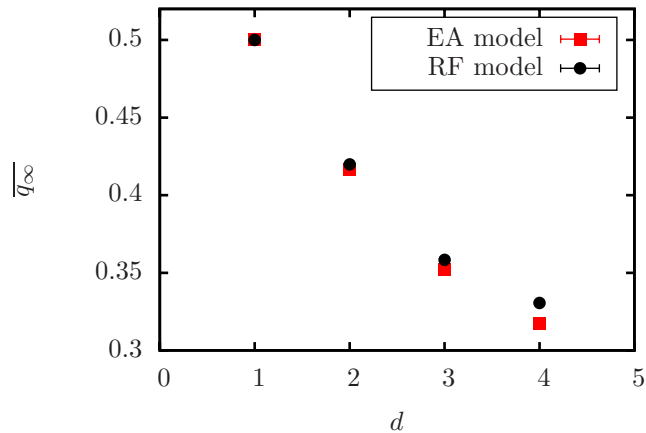}
\caption{The average overlap in the absorbing state for the largest systems size $\qfinalNmax$ vs.\ dimension $d$ for  Edwards-Anderson spin glass (red squares) and random ferromagnet (black circles) models. } 
\label{fig:QD} 
\end{figure}

The fast convergence of $\qfinalN$ to a constant value motivates using $\qfinalNmax$ as an estimator for $\qfinalinfN$ where $N_{\rm max}$ is the largest size simulated for each system.  For $d = 1$, it is known that $\qfinalinfN = 1/2$~\cite{NNS00}. The results for $\qfinalNmax$ and the values of $N_{\rm max}$ for each dimension are presented in Table \ref{tab:QD}. The errors listed in the Table include only statistical errors.  It is possible that systematic errors due to estimating $\qfinalinfN$ at a finite $N_{\rm max}$ are larger.
Figure \ref{fig:QD} shows $\qfinalNmax$ vs.\ $d$ and indicates that $\qfinalinfN$ decreases with increasing dimension  though with only four data points it is unclear whether in the infinite-dimensional limit $\qfinalinfN=0$ or $\qfinalinfN>0$.  
\begin{table}[H]
\centering{}
\caption{The largest simulated system size, $N_{\rm max}$ and $\qfinalNmax$ for the EA model in dimension $d$.}
\begin{tabular*}{0.35\textwidth}{@{\extracolsep{\fill}} c r r }
\hline
\hline
$d$ & $N_{\rm max}$ & $\qfinalNmax$ \\  
\hline
2 & $100^2$ & 0.4166(1)\\
\hline
3 & $50^3$ & 0.35233(3)\\
\hline 
4 & $20^4$ & 0.31746(3)\\
\hline
\hline
\end{tabular*}
\label{tab:QD}
\end{table}

\begin{figure}[H]
\centering 
\includegraphics[scale=1]{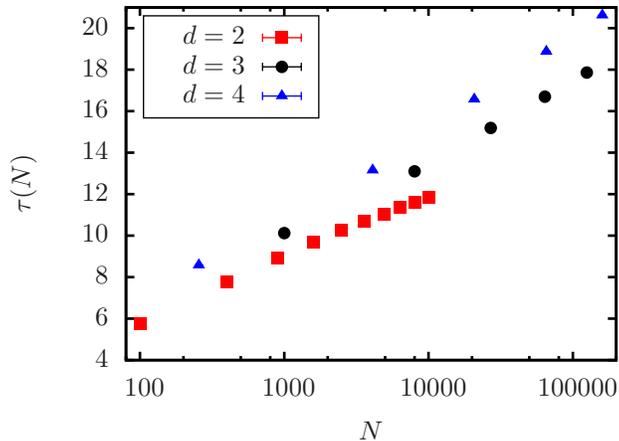}
\caption{(color online) Mean survival time $\mt(N)$ vs.\ number of spins $N$ for the EA spin glass models  $d=2$ (red squares),  $d=3$ (black circles), and $d=4$ (blue triangles). }
\label{fig:CT} 
\end{figure}

Next we consider the mean survival time $\mt(N)$ as a function of number of spins $N$.  The survival time for each system is the (integer) number of sweeps immediately prior to reaching the absorbing state. Figure \ref{fig:CT} shows $\mt(N)$  vs.~$N$ for each dimension $d$.
A slight downward curvature of $\mt(N)$ can be seen in two dimensions, but not in three and four dimensions, where $\mt(N)$  shows no sign of saturating up to the largest sizes studied. At the same time, we know from~\cite{NNS00} that in all dimensions the number of spin flips at each site is finite. It is also curious that in all cases the overlap saturates to a constant very quickly with $N$, but the mean survival time is still increasing, even for the largest sizes. How can we reconcile these disparities? In fact, these observations are not in contradiction. The most likely scenario is that $\mt(N)$ approaches a finite dimension-dependent constant for all finite $d$, and our simulations simply have not gone to sufficiently large $N$. But other possibilities are also consistent with observations. The fact that each spin flips finitely often of course does {\it not\/} necessarily imply that the survival time, or (roughly) equivalently, the mean number of spin flips per site, is finite: it could be the case that a small number of spins continue to flip long after most of the others have reached their final state. This would also be consistent with the overlap saturating while the survival time is still increasing with $N$. A more detailed numerical study is needed to determine whether this is indeed happening.

However, our main interest here is the dependence on $d$ of the mean number of spin flips per site at fixed $N$.  This is not the same quantity as $\mt(N)$, for the reasons discussed in the preceding paragraph; nevertheless, the two are related, and the clear increase in~$\mt(N)$ with dimension $d$ that can be seen in Fig.~\ref{fig:CT} provides reasonable supporting (more accurately, necessary but not sufficient) evidence that the typical number of spin flips per site increases with $d$, and we will take this as one of our conjectures in Sect.~\ref{sec:discussion}.

\begin{figure}[t]
\centering  
\includegraphics[scale=0.3]{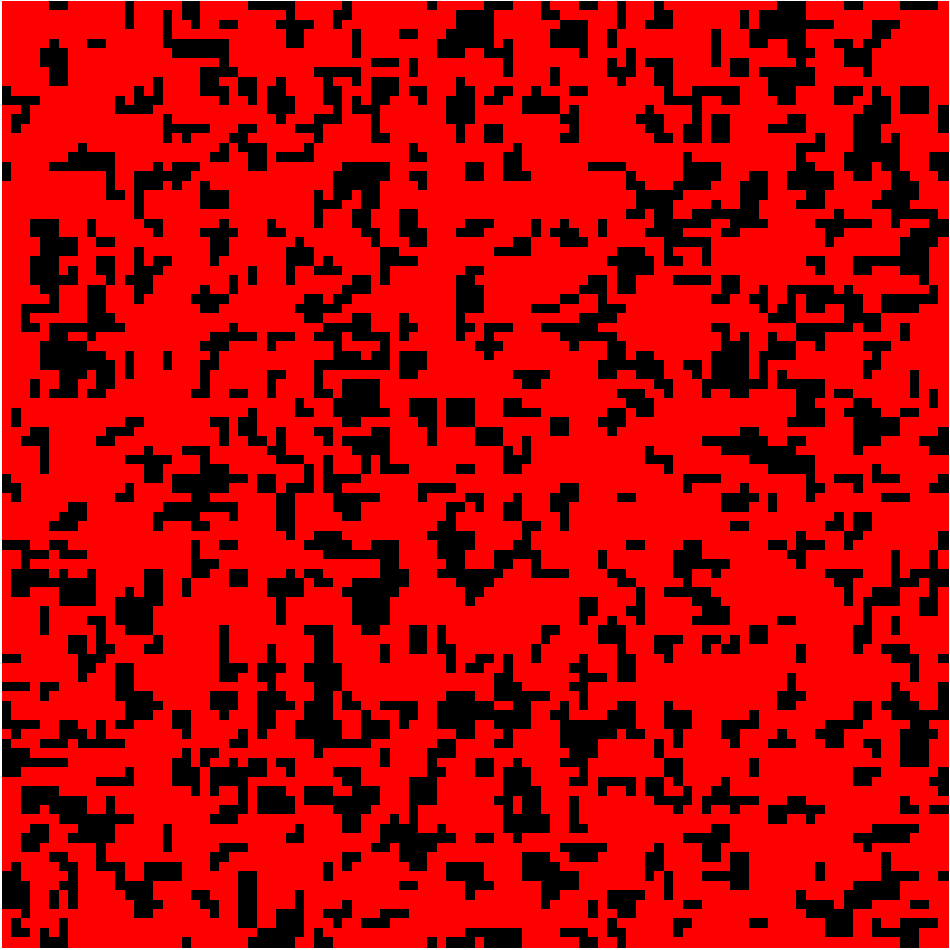}
\caption{(color online). The overlap configuration for a typical pair of final states for the two-dimensional Edwards-Anderson model with $L=100$.  Like spins percolate and are shown in red while unlike spins are shown in black.} 
\label{fig:config} 
\end{figure}

Finally, we consider the spatial structure of the overlap in the final state.  Figure \ref{fig:config} shows the overlap configuration for a typical pair of final states for the two-dimensional EA spin glass with $L=100$.  Like spins (shown in red) are well above the percolation threshold while unlike spins form small clusters.  Note that there are no isolated singleton unlike spins since those must be eliminated by the zero-temperature dynamics.

\subsection{Random ferromagnet in $d$ dimensions}
\label{subsec:RF}
\begin{figure}[t]
\centering 
\subfigure{ 
\label{fig:subfig:a}  
\includegraphics[scale=0.8]{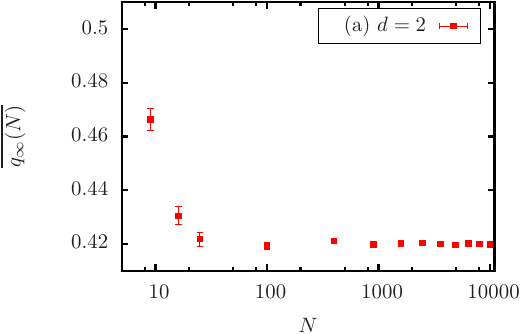}} 
 \subfigure{ 
\label{fig:subfig:b} 
\includegraphics[scale=0.8]{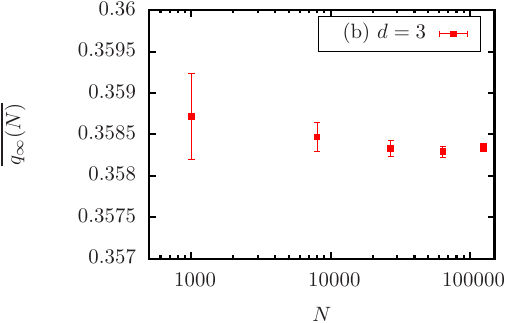}} 
\subfigure{ 
\label{fig:subfig:c} 
\includegraphics[scale=0.8]{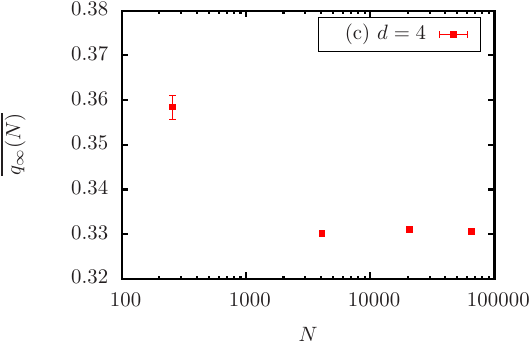}} 
\caption{The overlap $\qfinalN$ vs.\ system size $N$.  Panels (a), (b) and (c) represent the random ferromagnet in $d=2$, 3 and 4 dimensions, respectively.} 
\label{fig:QFD} 
\end{figure}

We now turn to studying similar questions for the random ferromagnet in $d$ dimensions.  Figure \ref{fig:QFD} is a plot of $\qfinalN$  as a function of number of spins $N$ for the random ferromagnet, with panels (a), (b), and (c) representing $d = 2$, 3, and 4 dimensions, respectively.  It is  clear that $\qfinalN$ is rapidly converging to a non-zero constant as $N \rightarrow \infty $.  

As in the case of the EA model, the fast convergence of $\qfinalN$ to a constant value  motivates  using $\qfinalNmax$ as an estimator for $\qfinalinfN$ where $N_{\rm max}$ is the largest size simulated for each system. The results for $\qfinalNmax$ and the values of $N_{\rm max}$ for each dimension are presented in Table \ref{tab:RQD}.  Figure~\ref{fig:QD} shows $\qfinalNmax$ vs.\ $d$ and indicates that $\qfinalinfN$ decreases with increasing dimension though with only four data points it is unclear whether in the infinite-dimensional limit $\qfinalinfN=0$ or $\qfinalinfN>0$.  The latter possibility is suggested by the finding (cf.~Sect.~\ref{subsubsec:mff}) that $\qfinalinfN=1$ for the random Curie-Weiss ferromagnet. This leaves two plausible scenarios: either $\qfinalinfN$ reaches a minimum at some finite dimension and then increases to one as $d\to\infty$, or else $\qfinalinfN\to 0$ as $d\to\infty$ and the infinite-dimensional limit is singular in the dynamics. In Sect.~\ref{subsec:sr} we argue that the latter possibility is the more likely of the two.

The results for the random ferromagnet are remarkably similar to those of the Edwards-Anderson model, especially in $d=2$  where $\qfinalNmax$ differ by only 0.3\%, though they differ by many standard errors.   In either case, the closeness of the two values of $\qfinalinfN$ suggest that frustration, which is present in the spin glass but not the random ferromagnet, plays little or no role in the nature vs.~nurture problem for low dimensionality.  As $d$ increases the difference between $\qfinalNmax$ for the EA model and RF model increases with the RF value larger than the EA value.  This difference becomes most stark when comparing the mean field models (see Sec.\ \ref{subsec:mf}).
 \begin{table}[H]
\centering{}
\caption{The largest simulated system size, $N_{\rm max}$ and $\qfinalNmax$ for the random ferromagnet in dimension $d$.}
\begin{tabular*}{0.35\textwidth}{@{\extracolsep{\fill}} c r r }
\hline
\hline
$d$ & $N_{\rm max}$ & $\qfinalNmax$ \\  
\hline
2 & $100^2$ & 0.4198(1)\\
\hline
3 & $50^3$ & 0.35835(5)\\
\hline 
4 & $16^4$ & 0.3305(1)\\
\hline
\hline
\end{tabular*}
\label{tab:RQD}
\end{table}

\begin{figure}[H]
\centering 
\includegraphics[scale=1]{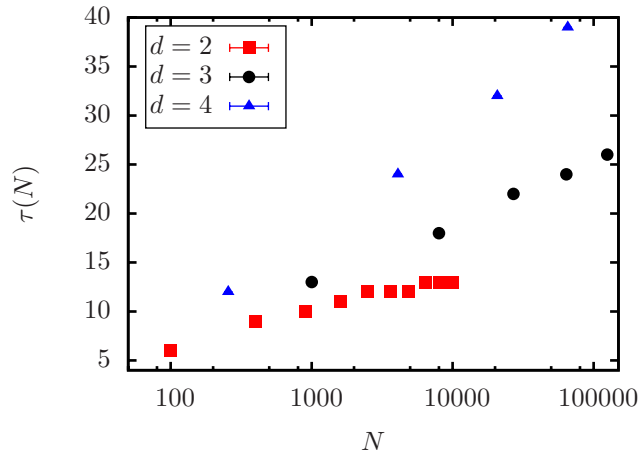}
\caption{(color online) Mean survival time $\mt(N)$ vs.\ number of spins $N$ for the random ferromagnet in  $d=2$ (red squares),  $d=3$ (black circles), and $d=4$ (blue triangles). }
\label{fig:CTr} 
\end{figure}

Next we consider the mean survival time $\mt(N)$ as a function of number of spins $N$.  The survival time for each system is the (integer) number of sweeps immediately prior to reaching the absorbing state. Figure \ref{fig:CTr} shows $\mt(N)$  vs.~$N$ for each dimension $d$. The apparent plateaus in the $d=2$ plots are due to measuring time as an integer number of sweeps. 

\subsection{Sherrington-Kirkpatrick model}
\label{sec:sk}
\begin{figure}[H]
\centering 
\includegraphics[scale=.4]{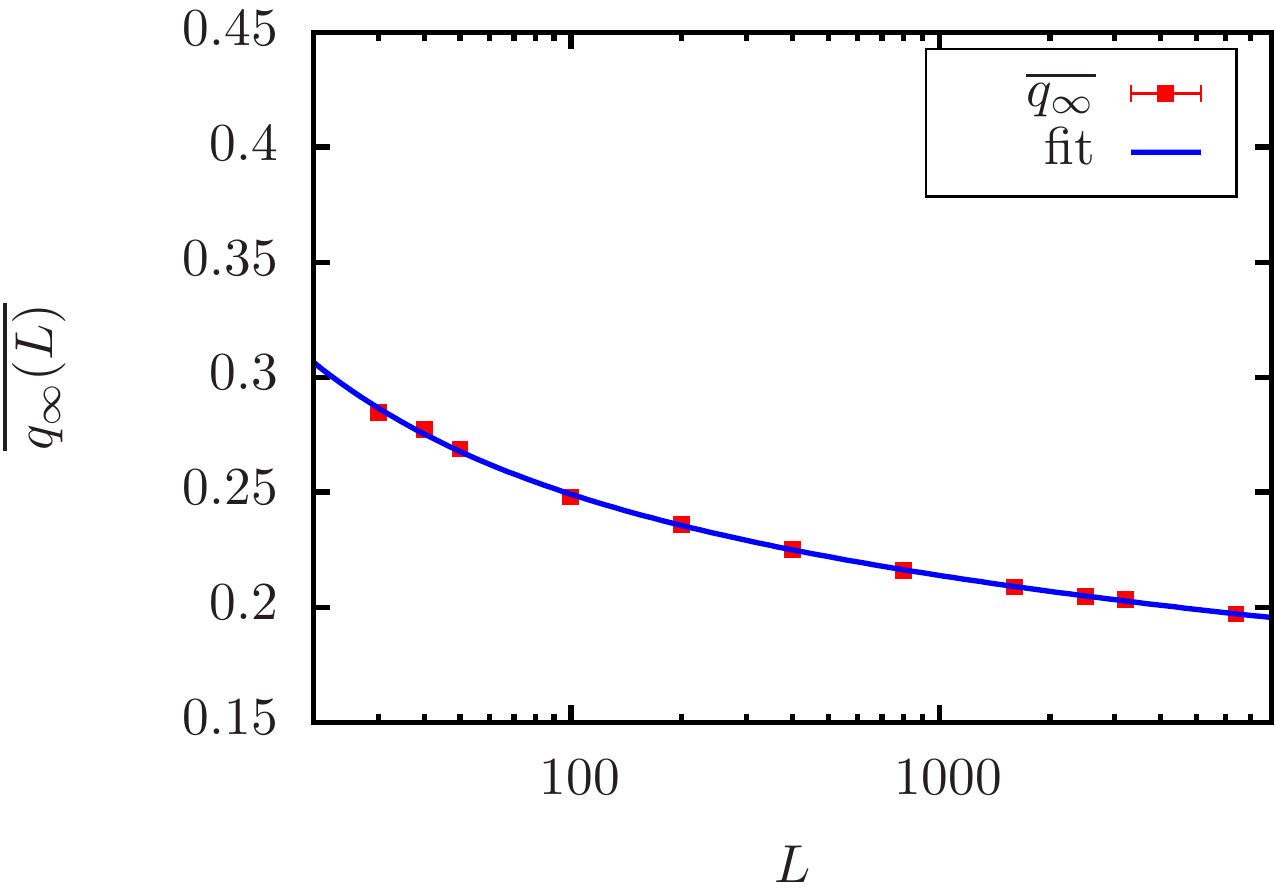}
\caption{The points are simulation results for $\qfinalN$ vs.\ $N$ for the SK model.  The curve is the highest quality fit, fit 1 with $N_{\rm min}=30$.} 
\label{fig:FIT} 
\end{figure}

This section presents  results for Sherrington-Kirkpatrick (SK) model, the Ising spin glass on the complete graph.
Figure \ref{fig:FIT} is a plot of $\qfinalN$  
as a function of $N$.  While it is clear that $\qfinalN$ is decreasing with $N$ it is not obvious whether $\qfinalinfN$ is zero or greater than zero.  
To attempt to determine which possibility holds we fit $\qfinalN$ to three functional forms:
\begin{enumerate}
\item $\qfinalN = \displaystyle\frac{a}{(\log{N})^{1/3}}+ \frac{b}{N}$\\
\item $\qfinalN = aN^c + b$\\
\item $\qfinalN = \displaystyle\frac{a}{(\log{N})^b}$
\end{enumerate}

For each of these forms, we carried out  fits for two values of the minimum size used in the fit $N_{\rm min}=30$ and 50. 
The fitting coefficients and quality of the fits are summarized in Table \ref{tab:FIT} and Fig.\ \ref{fig:FIT} shows the highest quality fit, which is fit 1 with $N_{\rm min}=30$.

\begin{table}[H]
\centering{}
\caption{Parameters and fit quality for the three fits of $\qfinalN$ vs. N}
\begin{tabular*}{0.9\textwidth}{@{\extracolsep{\fill}} |c | c c | c c | c c |}
\hline
\hline
& \multicolumn{2}{c|}{$\frac{a}{(\log{N})^{1/3}}+ \frac{b}{N}$}& \multicolumn{2}{c|}{$aN^c + b$}& \multicolumn{2}{c|}{$\frac{a}{(\log{N})^b}$}\\  
\hline
$N_{\rm min}$ & 30 & 50 & 30 & 50 & 30 & 50\\
\hline
a & 0.4063(3) & 0.4064(4) & 0.29(2) & 0.24(2) & 0.439(8) & 0.421(6) \\
\hline 
b & 0.49(5) & 0.41(13) & 0.175(3) & 0.166(6) & 0.369(9) & 0.349(7) \\
\hline
c & & & $-0.29(2)$ & $-0.23(3) $& &  \\
\hline
Reduced $\chi^2$ & 0.995 & 1.12 & 2.44 & 1.66 & 4.43 & 1.51 \\
\hline
Quality of Fit & 0.441 & 0.345 & 0.0123 & 0.141 & $7.94\times10^{-6}$ &0.170 \\
\hline
\hline
\end{tabular*}
\label{tab:FIT}
\end{table}
The first fitting function is the best fit to the data for both values of $N_{\rm min}$ and implies that $\qfinalinfN=0$.   Nonetheless, the second functional form also provides a reasonable fit and implies that $\qfinalinfN >0$.  It is also noteworthy that the leading coefficient~$a$ of the first functional form is stable with respect to changing the fitting range while this is not true of the second functional form.  Although the numerics are not definitive, we believe that it is most likely that $\qfinalN$ converges very slowly, i.e.\ as $1/\log(N)^{1/3}$, to $\qfinalinfN=0$.  Since, in some sense, the SK model is believed to correspond to the EA model at $d=\infty$, these results provide further support to the conjecture that $q_\infty\rightarrow 0$ in the EA~model as dimensionality goes to infinity.

\begin{figure}[H]
\centering  
\includegraphics[scale=1]{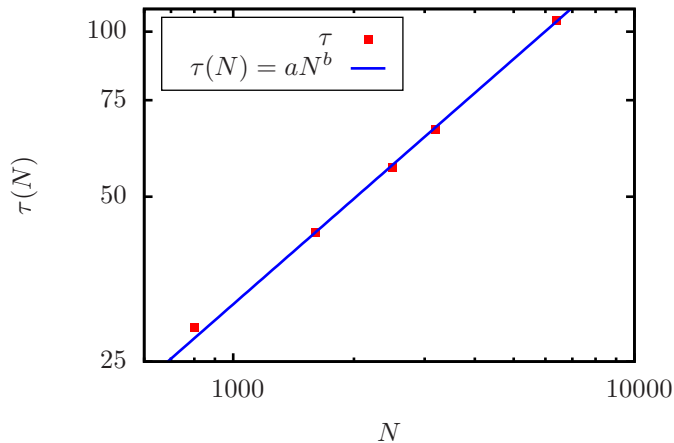}
\caption{The median time for reaching the absorbing state, $\mt(N)$ vs.\ number of spins, $N$ for the SK model.} 
\label{fig:CTN} 
\end{figure}

Next we consider the survival time as a function of $N$.
For the SK model, we define $\mt(N)$ as the median survival time for a system of $N$~spins.  We note that distribution of survival times is well described by a lognormal and obtain the median by fitting the data to a lognormal.   Figure \ref{fig:CTN} is a log-log plot of $\mt(N)$  vs.\ $N$.
A power law fit with a $1/N$ correction to scaling, $\mt(N) = a N^b (1 + c/N)$ does a reasonable job of fitting the data for $N \geq 800$.  The fitted exponent is  $b=0.69$.   A pure power law is a worse fit and yields a smaller value of the exponent, $b=0.64$. The pure power law with $b=0.64$ is shown as the solid curve in Fig. \ref{fig:CTN}.  Although we have low confidence in the value of the exponent describing $\mt$, in contrast to the finite-dimensional models, it seems clear that the survival time diverges in $N$, most likely as a power near $2/3$.

\begin{figure}[H]
\centering  
\includegraphics[scale=1.0]{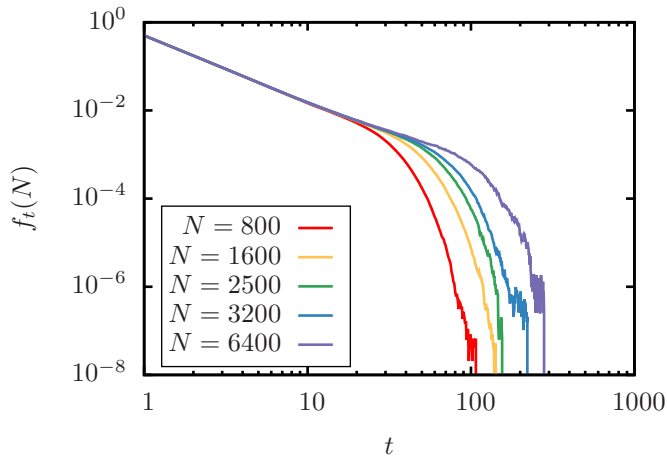}
\caption{(color online) Log-log plot of the fraction of active spins $\sizeFL$ vs.\ $t$ for the SK model for different sizes $N$ with sizes increasing from left to right. } 
\label{fig:FLS} 
\end{figure}

\begin{figure}[H]
\centering  
\includegraphics[scale=1.0]{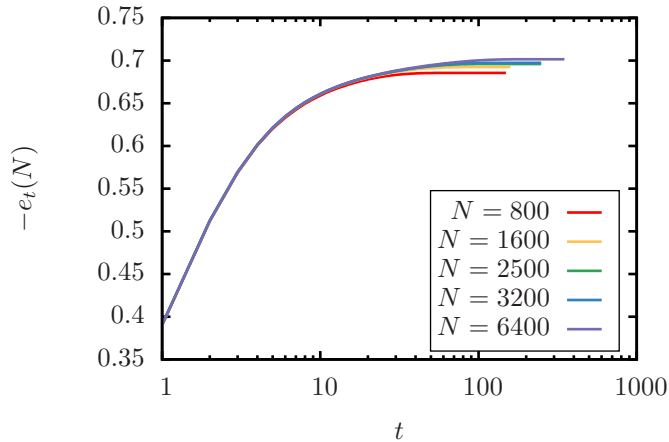}
\caption{(color online) The energy per spin for the SK model as a function of time with sizes increasing from bottom to top. }
\label{fig:SKET} 
\end{figure}

Next, we study the fraction of active (flippable) spins $\sizeFL$ as a function of time $t$. 
Figure \ref{fig:FLS} is a log-log plot of $\sizeFL$ vs.\ $t$. We find that for about the first 10 sweeps, $\sizeFL$ has a power law decay.   At longer times the curves fall much more steeply.  However, for intermediate times and the larger system sizes there appears to be a flattening before the steep fall-off.   Thus the asymptotic behavior in time for  $\sizeFL$ in the limit $N \rightarrow \infty$ is not clear.  A power law  $\sizeFL = (1/2)(t +1)^{-1.52}$ in the range $0 \leq t \leq 10$ is a good fit to the data for all values of $N$.  
Finally, we study the energy per spin.  Figure \ref{fig:SKET} shows the  energy per spin $e_t(N)$ for the SK model as a function of time $t$ for several system sizes $N$.  The energy per spin decreases rapidly and then reaches a plateau on the same time scale as the median time to absorption (see Fig.\ \ref{fig:CTN}).  For size $N=6400$, the mean energy per spin in the absorbing state is $-0.70156(2)$ whereas the mean ground state energy per spin for the SK model with $N=6400$ spins is about $-0.760$ \cite{Palassini08}.  The ground state energy for the infinite size SK model is $-0.76321$.  Thus, there are either quite large finite-size corrections or the energy resulting from a quench to zero temperature is significantly higher than the ground state energy.  The plateau values $e_\infty(N)$ can be reasonably fit as a function of $N$ to extrapolate either to the average ground state energy or to a value greater than the ground state energy.
\section{Discussion}
\label{sec:discussion}

In this section we discuss the numerical results presented above within the context of the theory of the nature vs.~nurture problem. We also examine
mean-field models for which rigorous conclusions can be obtained.

The numerical studies presented in this paper focus on three disordered systems: the EA~spin glass and random ferromagnet as a function of dimension~$d$, and the SK~spin glass as a function of the number of spins~$N$.
While our understanding (in the nature vs.~nurture context) of these models remains incomplete, the simplest and most natural extrapolation of the numerical results suggest several broad conclusions, which for now remain as conjectures:

\begin{enumerate}

\item While it can be rigorously shown (see below in Sect.~\ref{subsec:sr}) that the dynamical order parameter $q_\infty$ in both the EA~model and the random ferromagnet on $\mathbb Z^d$ lies strictly between zero and one for any finite $d$, extrapolation of numerical results indicates that both tend to zero as $d\to\infty$ (but see the remark at the end of the second paragraph of Sect.~\ref{subsec:RF}), and similarly for the SK~model as $N\to\infty$.  

\item In the numerical simulations of the EA model and random ferromagnet on~$\mathbb{Z}^d$ reported in Sect.~\ref{sec:results},
the survival time is seen to increase with dimension~$d$; based on this (see the discussion at the end of Sect.~\ref{subsec:EA}), we conjecture that the number of spin flips per site in both models diverges as $d\to\infty$.

\item  Above one dimension, $q_\infty$ is larger in the random ferromagnet than in the EA spin glass, and the difference between the two increases with dimension.

\end{enumerate}

We do not yet have rigorous proofs of these conjectures. We first present some analytical results for several mean-field models, and then use the resulting insights to develop some preliminary understanding of the dynamical behavior of short-range models.

\subsection{Mean-field models}
\label{subsec:mf}

The mean-field models discussed below have not previously been studied from the nature vs.~nurture perspective. Given that the numerical results for the SK model indicate that 
$q_\infty\to 0$ as $N\to\infty$, one might expect that in other mean field models $q_\infty\to 0$ as $N\to\infty$ as well. However, this is not the case. 

\subsubsection{The Random Energy Model}
\label{subsubsec:rem}

We begin by considering the random energy model~(REM)~\cite{Derrida80,Derrida81}. While the results presented below can be proved, we will present them here informally.  

Although the REM is not typically thought of as a dynamical model, there is a natural dynamics associated with it. 
For any $N$-spin system there are $2^N$ corresponding spin configurations, which are the corners of the hypercube. The REM assigns a random energy independently to each corner (or site) of the hypercube.
The distribution of the energies could be Gaussian, as in the original formulation of the REM, or flat, or some similar distribution; the results are independent of the specific form, as long as the distribution 
is continuous and its variance is finite. A local minimum then corresponds to a site whose $N$ neighboring sites 
on the hypercube all have larger energy. (This corresponds to a 1-spin flip stable state, because the spin configuration corresponding to each neighbor on the hypercube differs from that corresponding to the original site by a single spin flip.)

Consider now a random walk starting at an arbitrary site, corresponding to a uniformly chosen point on the hypercube. If the starting site is a local minimum the walk goes no farther.
Otherwise, if the site has $k$ (out of $N$) neighbors with lower energy, the walk chooses one uniformly at random among the $k$.
This is equivalent to the usual zero-temperature Glauber~dynamics on the Ising spin system formulation of the REM. One then continues the process until it ends at a local minimum.

We can then ask a number of well-posed questions; in particular, how long is the length of a typical walk? If such a walk has, on average,  $o(N)$ steps, i.e., does not proceed macroscopically far in the hypercube, then  $q_\infty$ must go to 1. 
But this is precisely the case for the REM. In fact, this problem was solved in a different context by Kauffman and Levin~\cite{KL87}, who found that, on average, such a walk takes only $O(\log N)$ steps. Consequently, $q_\infty\ge 1 - O(\log N/N)$, so as $N\to\infty$, $q_\infty\to 1$: nature always wins. This is an unusual result, which we have not seen for any nontrivial short-range model. (The largest $q_\infty$ found for these models is 1/2, for a random Ising chain in one dimension~\cite{NNS00}.) The REM is of interest because, despite its simplicity, it mimics much (though certainly not all) of the thermodynamics of the far more complex SK model. These arguments demonstrate, though, that the dynamics of the two models, at least from the viewpoint espoused in this paper, are very different.

\subsubsection{Mean-Field Ferromagnetic Models}
\label{subsubsec:mff}

We turn now to mean-field ferromagnetic models. In the case of the uniform Curie-Weiss model with $N$ spins, a typical initial condition will have an excess 
(of order $\sqrt{N}$) of spins in one state (say the plus state) over the other. Therefore, every spin feels (to $O(1/N)$) the same positive internal field, and this can only increase with time. Given the usual Glauber dynamics, it's clear that the final state will then be all plus, so the initial condition completely determines the final configuration. The only initial conditions in which this will not be true is that for which $-1\le\sum_{i=1}^NS_i\le 1$. It is easy to see that the contribution to $q_\infty$ from such configurations goes to zero as $N\to \infty$. The conclusion is that again $q_\infty\to 1$ as $N\to\infty$, but for different reasons: the number of spin flips in the ferromagnet is $O(N)$, rather than $O(\log N)$ as in the REM. For the REM, $q_\infty\to 1$ because there are numerous local minima, and the nearest one is typically $O(\log N)$ steps from a random starting configuration. For the ferromagnet, there are no metastable states: the system simply starts in one of the two available attractor basins, after which its future is completely determined. 

For disordered Curie-Weiss models, in which the couplings are i.i.d.\ nonnegative random variables, it can be shown that $q_\infty$ still goes to $1$ as $N\to \infty$, but more extensive arguments are required and a proof will appear in a future paper~\cite{GNSinprep}. Here we provide a brief sketch of the main physical ideas that underlie the mathematical proof. 

As with the uniform Curie-Weiss model, a typical initial spin configuration will have $O(\sqrt N)$ excess of plus or minus spins; as before, we consider the plus state. The difference with the uniform Curie-Weiss model is that now the internal fields acting on the spins will vary, and some will be negative. Consequently, one has to study the {\it distribution\/} of the internal fields at each site. It is easy to show that, at time zero, this distribution has positive mean. The main technical issue is then to show that the fraction of sites with positive internal field increases steadily with time: the proof demonstrates that after a time of order $N^{1/2+\epsilon}$, where $\epsilon > 0$ is independent of $N$, every site has positive internal field with probability going to one as $N\to\infty$. At this point, the dynamics monotonically leads to absorption into the all-plus state, so that $q_\infty\to 1$ as $N\to\infty$.

The proof demonstrates that the system (thought of as a random walker on the $2^N$ hypercube) rapidly descends down the energy landscape to one of the two uniform stable states.  This suggests a surprising possibility. If the random Curie-Weiss ferromagnet possesses many metastable states, like finite-dimensional random ferromagnets and spin glasses~\cite{NS99b} or the SK~model~\cite{BM80}, then the dynamics somehow avoids them. Of course, the REM, which has many metastable states, also has $q_\infty\to 1$; the difference is that the number of steps in the random walk to the final state is $O(N)$ in the random ferromagnet, while it is $O(\log N)$ in the REM, so the mechanism in the two cases is significantly different.  Because of the much larger distance (in state space) between the initial and final states in the random ferromagnet, one would expect that trapping in a metastable state would lead to $q_\infty<1$. The fact that this doesn't happen suggests that metastability is absent in the Curie-Weiss random ferromagnet (or else metastable states exist but have no effect on the dynamics, either being small in number and/or having small basins of attraction, and so are physically irrelevant).

\subsubsection{SK Model}
\label{subsubsec:SK}

The behavior of the SK~model is markedly different from the REM and the mean-field ferromagnets: here the numerics indicate that $q_\infty\to 0$ (albeit slowly) as $N\to\infty$. We do not at this point have a rigorous argument to support this conclusion, but heuristically it appears quite reasonable. At zero temperature, the SK model has exponentially many (in $N$) one-spin-flip metastable states~\cite{BM80}, so its dynamics differs from that of the mean-field ferromagnetic models. Moreover, the numerics show that the number of spin flips grows linearly with $N$, so its behavior also differs from the REM. Given the $O(N)$ distance traveled by the SK model on the state space hypercube to find an absorbing state, and given that these states have been shown to be uncorrelated~\cite{BM80}, the decay of $q_\infty$ as $N$ grows is not surprising.

\subsection{Short-range models}
\label{subsec:sr}

We begin with some relevant rigorous results obtained in~\cite{NS99b}.   These are the following: first, for both the EA~spin glass and the random ferromagnet on the infinite Euclidean lattice $\mathbb{Z}^d$ for any finite~$d$, there 
is an {\it uncountable\/} infinity of $k$-spin-flip stable states, for any finite $k\ge 1$ (states with $k=1$ comprise the set of final possible states for the dynamics described in this paper). The overlap distribution for the set of such states is a $\delta$-function at 0. 

Related to the above is a corresponding result, also proved in~\cite{NS99b}, that if one begins with two independently chosen random initial configurations for either the spin glass or random ferromagnet in any finite dimension, and lets their respective zero-temperature Glauber dynamics proceed independently, then the two dynamical runs will almost surely ``land'' in separate 1-spin-flip stable states with overlap zero. However, the nature vs.~nurture question focuses on a {\it single\/} randomly chosen initial configuration and asks for the spin overlap between the final states obtained through different runs under independent dynamical realizations. In~\cite{NS99b}, it was proved that with probability~one the spin overlap between final states obtained under two independent dynamical runs starting from the {\it same\/} initial configuration is none other than $q_\infty$.

We add here a further result, namely that for any finite $d$, $q_\infty$ is strictly between zero and one for both the random ferromagnet and the EA model with continuous coupling distributions on $\mathbb{Z}^d$. The proof behind this claim rests on the existence of so-called ``bully bonds'' --- that is, couplings that must be satisfied in any absorbing state. This concept was first presented in~\cite{NNS00}, and such bonds can be defined as follows: $J_{xy}$ is a bully bond with respect to $x$ if
\begin{equation}
\label{eq:bb}
|J_{xy}|>\sum_{z: |x-z|=1,z\ne y}|J_{xz}|\, .
\end{equation}
$J_{xy}$ is a bully bond with respect to $y$ if the analogous inequality with $x$ and $y$ exchanging roles is valid.

It is clear that such a coupling must be satisfied in ground states and all $1$-spin-flip stable states.  However, given that we're also interested in the dynamical process by which an initial state evolves to a final one, we add a further condition, namely that $J_{xy}$ is a bully bond with respect to {\it both\/} $x$ and $y$, i.e., for each of the two sites connected by the bond. The density of such ``double bully'' bonds decreases as $d$ increases, but is strictly positive for any $d<\infty$.

The contribution to $q_\infty$ from the set of all ``double bully'' bonds is always exactly 1/2 (in an infinite system), for the following reason: in a random initial configuration, half the bully bonds will be satisfied and half will be unsatisfied. Those that are satisfied in the initial spin configuration will remain satisfied for all time under any dynamical realization, giving $q_\infty > 0$. Those that are unsatisfied in the initial spin configuration, on the other hand, will contribute zero to $q_\infty$. This is because the final configuration of the two spins connected by the coupling can end up in one of two final, spin-reversed states, depending on which spin's Poisson clock rings first under a particular dynamical realization. Therefore, half of all dynamical realizations will result in one orientation, and the other half will result in the opposite orientation, giving a contribution to $q_\infty$ of zero. Therefore $q_\infty < 1$.

With these preliminary remarks, we now turn to a discussion of the two short-range models under consideration.

Evidence for the conjecture that $q_\infty\to 0$ as $d\to\infty$ in finite-dimensional EA~spin glasses is strengthened by the numerical and theoretical results presented above: in the $N$-spin SK~model, $q_\infty\to 0$ as $N\to\infty$, and we expect this to describe the dynamical behavior of the EA spin glass in the limit of infinite dimensionality. However, the same behavior appears absent for the random ferromagnet: its infinite-range version has $q_\infty\to 1$. One of two things then presumably occurs: either (as the numerics indicate) $q_\infty$ for the random ferromagnet on $\mathbb{Z}^d$ goes to zero monotonically as $d\to\infty$ --- in which the case the $d\to\infty$ limit is singular in the dynamics --- or else $q_\infty$ begins to increase again above some dimension. The latter sounds implausible but cannot be ruled out at this stage.

Nevertheless, we suspect that the first possibility holds, and that the $d\to\infty$ limit is indeed singular in the dynamics of the random ferromagnet. We present a brief heuristic argument here to support this claim. 

We note first that amongst connected graphs, the $1D$ chain and the complete graph are extremal. With respect to the nature vs.~nurture question, in the $1D$ case {\it every\/} spin lives in a domain whose final state is completely determined by the final orientation of the spins at the two endpoints of a double bully bond~\cite{NNS00}. In sharp contrast, the disorder in infinite-range random ferromagnets plays no role in determining the final state: it is completely determined by the initial bias. In all other models, there is an interplay between the two kinds of disorder present: the coupling disorder and the randomness in choosing the initial spin configuration. In one dimension, the interplay between the two leads to a $q_\infty$ of exactly 1/2: half of the double bullies are satisfied in the initial configuration, which then determines the state of all spins lying in their domain; and half are unsatisfied, giving (as in the discussion above) a contribution of zero to $q_\infty$. So in one dimension the double bullies determine everything, but as dimension increases, their effect in determining the final state diminishes --- an increasing fraction of bonds don't lie in the region of influence of a double bully, and more extensive arguments can rule out more complicated influence domains resulting from the disorder as $d\to\infty$. Effectively, the quenched disorder is playing a decreasing role in determining dynamical outcomes.

We now consider the internal effective field, due to neighboring spins, at each site.  Consider a fixed random initial configuration, and choose two neighboring sites $i$ and $j$. We consider the internal fields at these two sites as random variables under the law of the coupling disorder. In the random Curie-Weiss ferromagnet, the correlation between the two can be shown to be $1-O(1/N)$, while in the short-range EA model and random ferromagnet the correlation decays as $1/d$.  In the mean-field case, the strong correlation at time zero leads to the system magnetization undergoing a biased random walk, so that the internal field at each site increases in time. In contrast, the high-dimensional random ferromagnet initially executes an (almost) unbiased random walk, at least at early times, due to the lack of correlations. As a result, internal fields at different sites are (roughly) as likely to end up positive as negative.
We noted earlier that the effect of quenched disorder is likely to decrease as dimension increases; the preceding argument indicates that the effect of the random initial configuration also decreases. All that is left is the dynamics, so one would expect $q_\infty$ to decrease as well. Whether this rough argument can be refined and be made at least partly rigorous is under investigation.

What seems clear, though, is that two conditions appear to be necessary in order that $q_\infty < 1$. The first is the presence of a large number of uncorrelated metastable states, so that the system has many possible final states whose overlap is small. This condition is satisfied for the EA model and random ferromagnet in all finite dimensions, as well as by the SK model and the REM.  The second, equally important, condition is that a finite system undergoes at least $O(N)$ spin flips before reaching the absorbing state, and correspondingly for an infinite system, the average number of spin flips per site is strictly positive. This is the case for the EA model and random ferromagnet in all finite dimensions, as well as by the SK model and both mean-field ferromagnets. The systems satisfying both properties are the SK model and the EA model and random ferromagnet in all finite dimensions, and in all of these the numerics suggest that $q_\infty\to 0$ in the appropriate limit.

\begin{acknowledgements} 

JM was supported in part by NSF DMR-1208046 and DMR-1507506. RG, CMN and DLS were supported in part by
NSF DMS-1207678.  DLS thanks the Guggenheim Foundation for a fellowship that partially supported this research. Simulations were performed on the Courant
Institute of Mathematical Sciences computer cluster. JM and DLS thank the Aspen Center for Physics (under NSF Grant 1066293), where some of
this work was done.
\end{acknowledgements}

\small\def\em{\it} \newcommand{\noopsort}[1]{} \newcommand{\printfirst}[2]{#1}
  \newcommand{\singleletter}[1]{#1} \newcommand{\switchargs}[2]{#2#1}


\begin{thebibliography}{10}
\bibitem{note2d} Although quantities introduced in this section, such as the heritability exponent and the spin overlap, are discussed
within the context of the two-dimensional uniform ferromagnet, their definitions are valid (suitably generalized) in any dimension.
\bibitem{notezero} It is possible in principle for $q_\infty=0$ for a system where every spin flips only finitely many times. However, in all
of the nontrivial physical models studied so far in which $q_\infty=0$, $q_t$ approaches zero asymptotically as $t\to\infty$, and we expect this to be the case
more generally.
\end{thebibliography}
\end{document}